\documentclass[]{aa}

\usepackage[varg]{txfonts}

\usepackage{lineno}
\usepackage{natbib}
\usepackage{amsmath}
\usepackage{graphicx}
\usepackage{wasysym}
\usepackage{txfonts}
\usepackage{xspace}
\usepackage{url}
\usepackage{textcomp}
\usepackage{multirow}
\usepackage{lipsum}

\newcommand{\ngculx}{NGC\,5907~ULX\xspace}

\newcommand{\pth}{NGC\,7793~P13\xspace}

\newcommand{\swift}{\textsl{Swift}\xspace}

\newcommand{\chandra}{\textsl{Chandra}\xspace}
\newcommand{\xmm}{\textsl{XMM-Newton}\xspace}

\newcommand{\nustar}{\textsl{NuSTAR}\xspace}

\def\nustar{{\it NuSTAR}\xspace}

\newcommand{\asec}{\ensuremath{''}\xspace}

\newcommand{\snr}{S/N\xspace}

\newcommand{\msun}{\ensuremath{\text{M}_{\odot}}\xspace}
\newcommand{\rsun}{\ensuremath{\text{R}_{\odot}}\xspace}

\newcommand{\ergcms}{\ensuremath{\text{erg\,cm}^{-2}\text{s}^{-1}}\xspace}
\newcommand{\ergps}{\ensuremath{\text{erg\,s}^{-1}}\xspace}
\newcommand{\heii}{\ensuremath{\rm He\,{\small II}}}

\begin{document}

\title{A tale of two periods:  determination of the orbital ephemeris of the super-Eddington pulsar NGC\,7793 P13}

\author{F.~F\"urst\inst{1}\and D.~J.~Walton\inst{2}\and M.~Heida\inst{3}\and F.~A.~Harrison\inst{3}\and D.~Barret\inst{4,5}\and M.~Brightman\inst{3}\and A.~C.~Fabian\inst{2}\and M.~J.~Middleton\inst{6}\and C.~Pinto\inst{2}\and V.~Rana\inst{7}\and F.~Tramper\inst{1}\and N.~Webb\inst{4,5}\and P.~Kretschmar\inst{1}}

\institute{European Space Astronomy Centre (ESAC), Science Operations Departement, 28692 Villanueva de la Ca\~nada, Madrid, Spain
\and Institute of Astronomy, Madingley Road, Cambridge CB3 0HA, UK
\and Cahill Center for Astronomy and Astrophysics, California Institute of Technology, Pasadena, CA 91125, USA
\and Universit\'e de Toulouse; UPS-OMP; IRAP; Toulouse, France
\and CNRS; IRAP; 9 Av. colonel Roche, BP 44346, F-31028 Toulouse cedex 4, France
\and Department of Physics and Astronomy, University of Southampton, Highfield, Southampton SO17 1BJ, UK
\and Raman Research Institute, C. V. Raman Avenue, Sadashivanagar, Bangalore - 560080, India}

\abstract{
We present a timing analysis of multiple \xmm and \nustar observations of  the ultra-luminous pulsar NGC\,7793 P13 spread over  its 65\,d variability period. We use the measured pulse periods to determine the orbital ephemeris, confirm a long orbital period with  $P_\text{orb}=63.9^{+0.5}_{-0.6}$\,d, and find an eccentricity of $e\leq0.15$. The orbital signature is imprinted on top of a secular spin-up, which seems to get faster as the source becomes  brighter. We also analyse data from dense monitoring of the source with \swift and find an optical photometric period of $63.9\pm0.5$\,d and an X-ray flux period of $66.8\pm0.4$\,d. The optical period is consistent with the orbital period, while the X-ray flux period is significantly longer. We discuss possible reasons for this discrepancy, which could be due to a super-orbital period caused by a precessing accretion disk or an orbital resonance.  
We put the  orbital period of P13 into context with the orbital periods implied for two other ultra-luminous pulsars, M82~X-2 and NGC\,5907~ULX and discuss possible implications for the system parameters.
}


\date{Received 24.04.2018 / Accepted 13.05.2018}

\maketitle

\section{Introduction}
\label{sec:intro}

Ultra-luminous X-ray pulsars (ULPs) are accreting neutron stars which show clearly detected pulsations and have inferred isotropic luminosities above $L >10^{39}\,\ergps$. These very high luminosities make them clearly super-Eddington and therefore a challenge to our  understanding of accretion. Currently, four ULPs are known, all located in nearby galaxies: M82~X-2 \citep{bachetti14a}, \ngculx \citep{israel17a}, NGC\,300~ULX1 \citep[aka SN2010da,][]{carpano18a}, and \pth \citep[hereafter P13,][]{p13, israel17b}. \citet{brightman18a} also identified  the ultra-luminous source M51~ULX\,8 as a likely neutron star accretor through the discovery of a cyclotron resonant scattering feature. However, no pulsations have been discovered from this source to date.

Before the discovery of pulsations, it was generally assumed that the compact object in  ultraluminous X-ray sources (ULXs) was a black hole, as the observed luminosities are more readily explained with a more massive compact object. To reach these luminosities in a sub-Eddington disk accretion regime, masses of $10^{2-5}$\,\msun were implied \citep[e.g.,][]{miller04b}. However, spectroscopic evidence, first by \xmm \citep[e.g.][]{stobbart06a, gladstone09a} and later in particular by \mbox{\nustar}, has shown that for most ULXs  stellar-remnant compact objects accreting at super-Eddington luminosities are more likely \citep[e.g.,][]{bachetti13a, walton13a, walton14a, walton15a, walton15b, rana15a, mukherjee15a, ngc5907, walton18b, walton18a}. 
For \pth, \citep{motch14a} could place a stringent constraint on the dynamical mass of the accretor of $< 15$\,\msun, confirming the super-Eddington explanation (of course, the later discovery of pulsations from P13 limits the maximum mass even further).
The spectral similarity between ULPs and bright ULXs opens the possibility that most known ULXs are actually powered by an accreting neutron star, and that the expected pulsations are suppressed or diluted through a yet to be identified process \citep{pintore17a, walton18b}.

Due to their large distances and sometimes high extinction, the optical counterparts of ULXs are difficult to identify. P13 is an exception, initially discovered in the X-rays by \citet{read99a}, for which the mass donor has been identified as a B9Ia super-giant \citep{motch11a}. \citet{motch14a} identified an optical and UV photometric period of $\approx$64\,d, which is also present in the radial velocity of the \heii\ emission. While the origin of the \heii\ emission line is debated \citep{fabrika15a}, \citet{motch14a} interpret the clearly detected period as  the orbital period  of the system and find a dynamical mass constraint of  $<$15\,\msun.

A  similar but significantly different period was  identified in  X-rays by \citet{hu17a} at $65.05\pm0.1$\,d, using long-term \swift/XRT monitoring data. \citet{hu17a} discuss the difference between the optical/UV and X-ray period, and propose that it might be due to a beat frequency with a super-orbital period. The super-orbital period could be caused by a warped and precessing accretion disk \citep{hu17a}. The arrival times of the optical maxima have been observed to vary, indicating the presence of a very long ($\sim$2500\,d) super-orbital period 
\citep{motch14a, hu17a}. 

On the other hand, \citet{hu17a} also discuss the possibility that both $\sim$65\,d periods could actually be super-orbital in nature, and that the true orbital period is much shorter. This argument is based on the fact that the  orbital periods for  two other known ULPs are only on the order of a few days: 2.5\,d for M82~X-2 \citep{bachetti14a} and likely $\sim$5\,d for \ngculx \citep[although in this case longer periods up to a few months could  not be ruled out at the 3-$\sigma$ level;][]{israel17a}. The observed phase jitter in the optical period in P13 could furthermore indicate an only semi-periodic behavior, which is expected for super-orbital periods but unlikely for orbital periods \citep{p13}. \citet{middleton18a} suggest that this period is due to Lense-Thirring precession of the accretion flow, which might provide constraints on the equation-of-state of the neutron star.

Identifying  the correct orbital period is an important step in understanding these systems, in particular to inform formation and evolution models \citep{marchant17a}. The best way to directly determine the orbital ephemeris in the X-ray band is to use the variation of the pulse period as function of orbital phase. For geometries where we do not view the system face-on, the pulse period will vary periodically due to the Doppler effect. 

With a pulse period of $P\approx415$\,ms, P13 shows the fastest spin of all known ULPs, making it an ideal candidate for this search. The spin-period can be determined to within 10\,ns with \xmm with exposure times around 50\,ks. The secular spin up $\dot P \approx 3.5\times10^{-11}$\,s\,s$^{-1}$ has been well established over a 3 year base-line \citep[2013--2016, ][]{p13,israel17a}.

To measure the influence of the orbital motion on $P$ and $\dot P$, and thereby to determine the ephemeris, we obtained five \xmm and three \nustar observations, spread out semi-regularly over 65\,d. This allows us to determine the pulse period at different phases. Those observations were supported by further \xmm and \nustar observations, which increase the baseline to determine the secular spin-up trend.  Table~\ref{tab:perevol} gives an overview of the data used.

Here we present the timing analysis of those data, while the spectral analysis will be presented in a forthcoming work. In Sect.~\ref{sec:data} we describe the observations and our data reduction. In Sect.~\ref{sec:analysis} we briefly describe the analysis method, present the timing results, and describe them with a model for the orbital variation. In Sect.~\ref{sec:disc} we discuss the implications of our results.
Uncertainties are given at the 90\% confidence level, unless otherwise noted.

\section{Observations and Data Reduction}
\label{sec:data}

\subsection{\swift}

\pth has been frequently monitored by the \textsl{Neil Gehrels Swift Observatory} \citep[\swift;][]{swiftref}  across several periods, first
for about $\sim$10 weeks in 2010, then for $\sim$6 months spanning late 2012 and
early 2013, for another $\sim$5 weeks spanning late 2014 and early 2015, and then
more-or-less persistently since May 2016. Although there are occasionally longer
gaps even within these monitored windows, the typical observing cadence is 5--10
days, and the average exposure is $\sim$2\,ks. Figure
\ref{fig:lc_uvot_xrt} shows the long-term light-curve of these observations obtained with the XRT and UVOT instruments
(\citealt{swiftxrtref, swiftuvotref}). 

We extracted all available XRT data between 2010-08-16 (ObsID 00031791001) and 2018-01-25 (ObsID 00093149030) with the standard \swift/XRT processing pipeline \citep{evans09a}, adding over 50 more observations to the light-curve compared to the data presented by \citet{hu17a}. The data are binned such that there is a single 0.3--10\,keV flux
measurement  from each observation. Aside from the 2012--2013 data, during which the
source was in a low-flux state and was not detected within any individual
observation \citep{motch14a, p13}, an indication of a periodicity on the
order of $\sim$65 days is visibly present in the X-ray data, with the observed count rate varying by
a factor of $\sim$3--4 from peak to trough.

To extract the UVOT \citep{swiftuvotref} data we downloaded all available \swift data of NGC 7793. We used a circular source region with a radius of 5\asec, as recommend by the \swift UVOT Software Guide\footnote{v2.2, \url{https://swift.gsfc.nasa.gov/analysis/UVOT\_swguide\_v2\_2.pdf}} centred on the \chandra coordinates for P13 \citep[RA = 23h 57' 50.9", Dec = $-32^\circ$37'26.6"; ][]{pannuti11a}. As background region we chose a circular region with a radius of 15\asec, located 35\asec east of P13 in a source-free region. As the source is located in the outskirts of the galaxy, some contamination of the magnitudes from the galactic background light is present. However, as we are only studying the variability of the source here, this should not influence our results. The UVOT data were extracted using the standard software as distributed with HEASOFT v6.20, in particular we used \texttt{uvotimsum} to sum up all exposures in each ObsID and \texttt{uvotsource} to extract the source magnitude.

\subsection{NuSTAR}

Data from \nustar \citep{harrison13a}  were reduced using the standard pipeline, \texttt{nupipeline}, provided
in the \nustar Data Analysis Software (v1.8.0), with standard filtering and \nustar\ CALDB v20171204.
Source products were extracted from
circular regions of radius 70$''$ for both focal plane modules (FPMA/B) using
\texttt{nuprodcuts}, with the background measured from circular regions with a radius of 120$''$ on the same
detectors as P13.
Light-curves were extracted in the 3--78\,keV energy band with a maximal time resolution of 0.1\,s.
All time information was transferred to the solar barycentre using the DE-200 solar system ephemeris \citep{solarDE200}.

\subsection{XMM-Newton}

Data from \xmm \citep{xmmref}  were reduced with the \xmm Science Analysis System (v15.0.0),
following the standard prescription.\footnote{http://xmm.esac.esa.int/} Owing to its
superior time resolution of 73.4\,ms, in this work we only consider data from the EPIC-pn detector
\citep{pnref}. The  data were taken in full frame mode and raw data  files were cleaned and calibrated using \texttt{epchain}. 
Source products for the 2016 data were extracted from circular regions of radius $\approx$40$''$, following the same method as described in \citet{p13}. For epochs 2017A--2017G we used a smaller circular region with a radius of $20''$, to avoid contamination from another transient source close to P13. For the latest epochs we used a circular region with a radius of $30''$, as the other source had faded. Background flaring was only problematic in epoch 2017A, for all other observations the full exposure time could be used (see Sect.~\ref{susec:puls}).
See Table~\ref{tab:perevol} for a complete observation log.

\section{Analysis}
\label{sec:analysis}

\begin{figure*}
\centering
\includegraphics[width=0.95\textwidth]{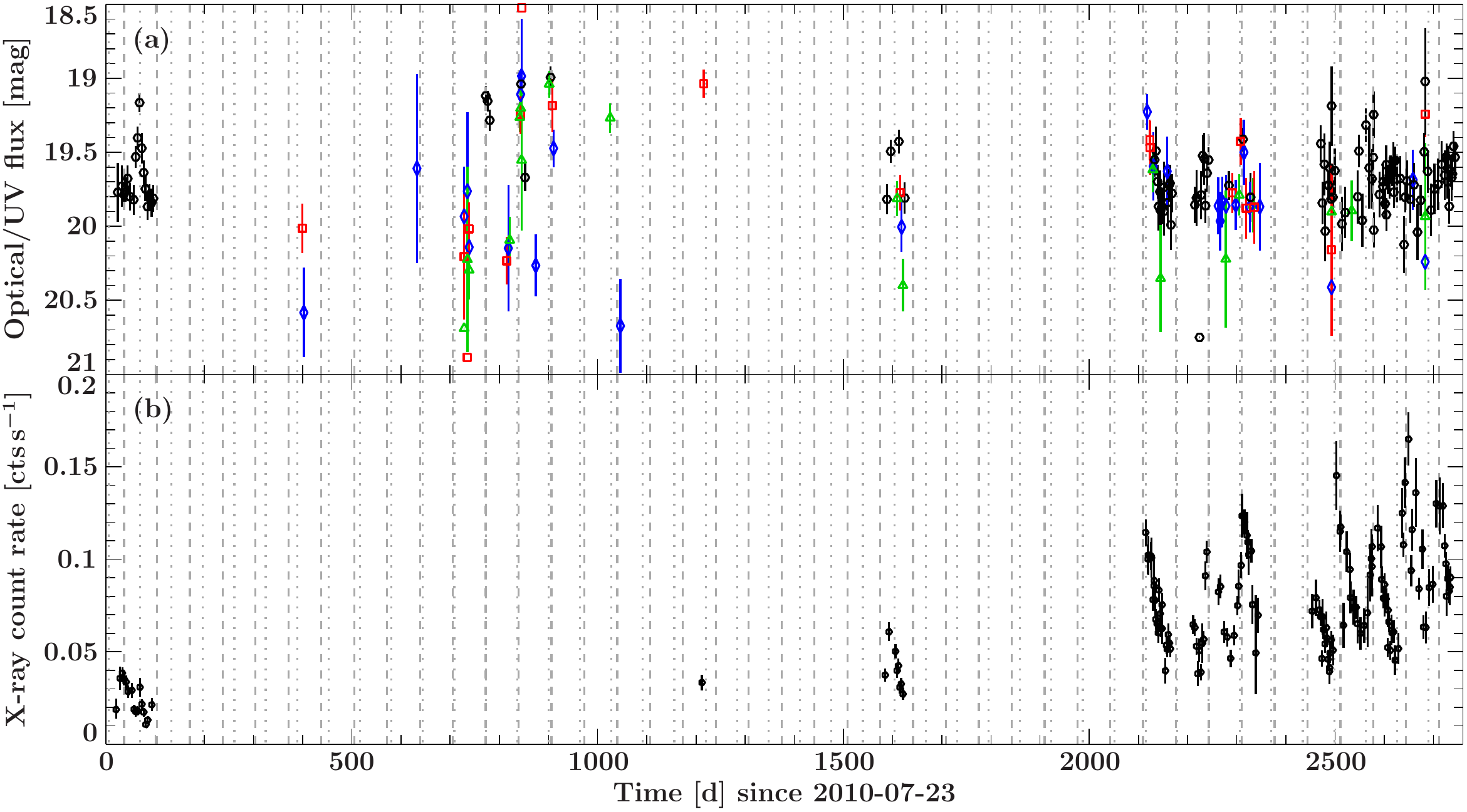}
\caption{\textit{(a)} Light-curve from \swift/UVOT, with one datapoint per observation. Black circles are u-filter,  red squares are uvw1-filter, green triangles are uvw2-filter, and blue diamonds are uvm2-filter. \textit{(b)} \swift/XRT light-curve in the 0.3--10\,keV energy band, with the same binning as the UVOT light-curve. Dotted lines mark the times of maximum light in the UVOT data, assuming a constant period of $P_\text{opt} = 63.9$\,d and dashed lines mark the maximum of the folded XRT profile with a constant period of $P_\text{X} = 66.9$\,d.
}
\label{fig:lc_uvot_xrt}
\end{figure*}

\subsection{Optical and UV}
\label{susec:optical}
We first perform a timing analysis of the UVOT data to confirm and update the optical period discovered by \citet{motch14a}, who used mainly data from ground-based telescopes. These authors found an optical period between $\sim$63.5--65.3\,d, depending on the exact data-set used, with considerable phase jitter for the time of maximum light. Their best-fit solution implies a period of 63.52\,d, with a super-orbital period of 2620\,d, on which the phases of maximum light move around the average. 
\citet{hu17a} used \swift/UVOT data up to December 2016 and found a period of $64.24\pm0.13$\,d.

Most UVOT observations were performed with the ``u'' filter (central wavelength 3465\,\AA, 114 observations), but some were also performed using the ``uvw1'' (2600\,\AA, 19 observations), ``uvw2'' (1928\,\AA, 20 observations), and ``uvm2'' (2246\,\AA, 27 observations) filter. See \citet{poole08a} for details about the UVOT photometric filters. The observed magnitudes are comparable between all filters, only the ``uvm2'' filter measures slightly lower values, as shown in the top panel of Fig~\ref{fig:lc_uvot_xrt}. Nonetheless, we decided to use all data together to search for periodicities, to obtain the best signal statistically. Using only the ``u'' filter data gives consistent results, albeit with larger uncertainties.

We search for periodicities using a Lomb-Scargle periodogram \citep{scargle82a} as well as epoch-folding \citep{leahy83a}. 
To determine the uncertainties on the  period, we simulate 5000 light-curves based on the folded profile at the strongest period, with an added white noise term to obtain the same variance as in the original data \citep{davies90a}. For each simulated light-curve we perform the same period search as for the real data, measure the strongest period, and report the 1-$\sigma$ standard deviation of the distribution of these periods.

\begin{figure}
\centering
\includegraphics[width=0.95\columnwidth]{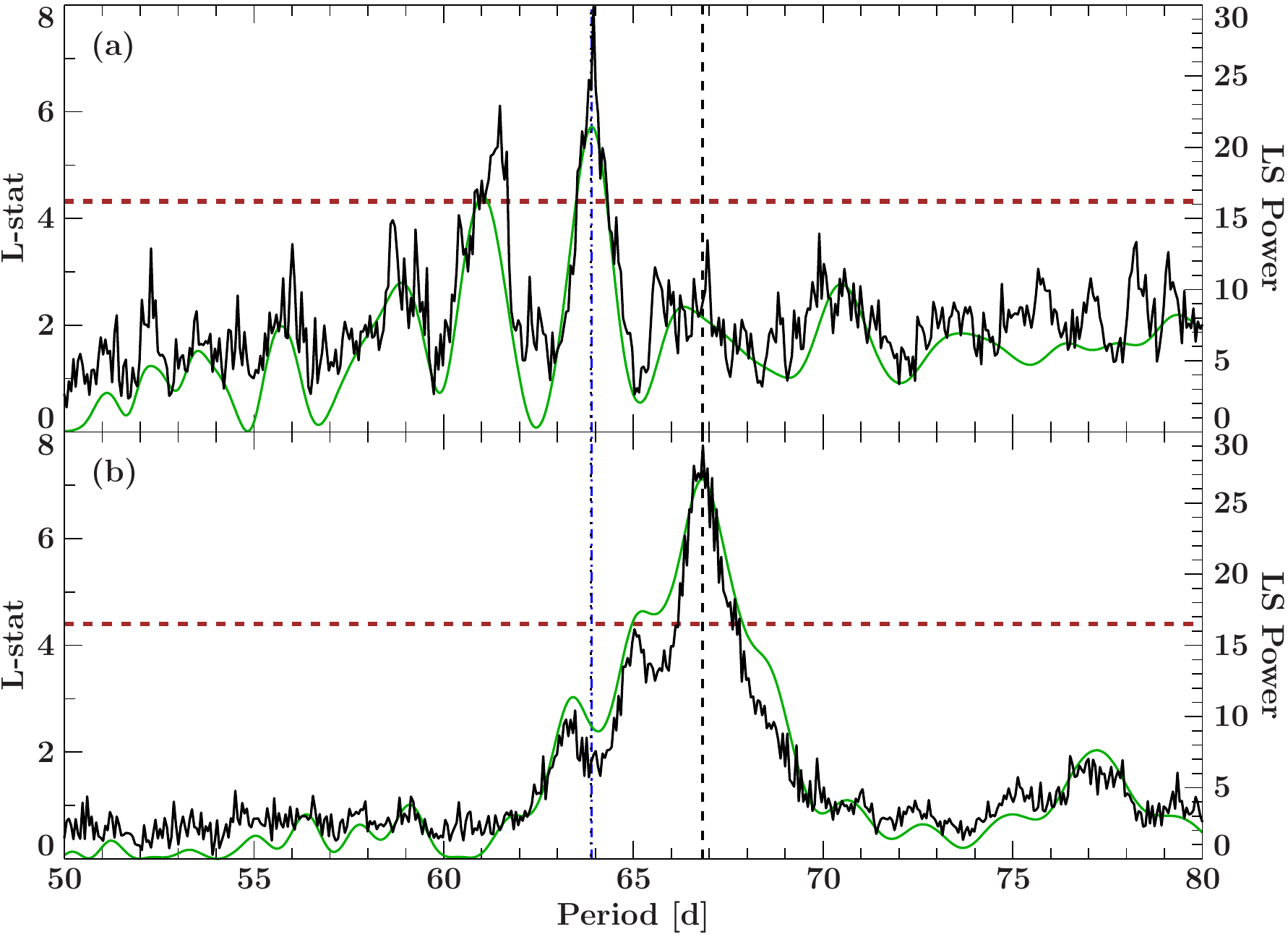}
\caption{\textit{(a)} Epoch-folding (black, left $y$-axis) and Lomb-Scargle Periodogram (green, right $y$-axis) of all \swift/UVOT data. \textit{(b)} Same as \textit{(a)} but for the \swift/XRT data. The dotted vertical line marks the strongest period of the UVOT data (which is consistent with the implied orbital period from the pulse period analysis), the dashed line the one of the XRT data. The dashed horizontal line marks the 99.9\% false alarm probability of the epoch folding results based on an F-distribution corrected for number of trial periods \citep{davies90a}.
}
\label{fig:uvot_xrt_epf}
\end{figure}

We find a strong signal at $63.9\pm 0.5$\,d, very well in agreement with the results by \citet{motch14a} and \citet{hu17a}, as shown in the top panel Fig.~\ref{fig:uvot_xrt_epf}. We fold the UVOT data on that period, and mark the times of maximum light based on the folded profile in Fig.~\ref{fig:lc_uvot_xrt}. As can be seen, in the data after 2016 April ($t=2100$\,d) the times of maximum light are difficult to identify in the UVOT light-curve. We therefore refrain from trying to fit a jitter of arrival times with a super-orbital period, as has been done by \citet{motch14a} and \citet{hu17a}.

\subsection{X-rays}
\label{susec:xrays}

As shown in the bottom panel of Fig.~\ref{fig:lc_uvot_xrt}, the X-rays also show strong variability on a similar time-scale as the optical data. \citet{hu17a} identified an X-ray period of  $65.05\pm0.1$\,d after subtracting the average count-rate of each observational epoch. This method of subtracting the average count-rate is used to make the fainter observations in 2010 better comparable to the observations in 2015--2017.  To measure the periodicity, we follow a slightly different approach here, and remove a linear brightening trend from the data. This approach is motivated by the fact that the average flux of P13 seems to be continuously brightening in the X-rays since the beginning of the observations. We also tried using a quadratic trend, but found no significant differences in the timing result.

Similar to our search in the UVOT data, we calculated the Lomb-Scargle periodogram and the epoch folding power using the detrended XRT light-curve. The results are shown in the bottom panel of Fig.~\ref{fig:uvot_xrt_epf} and show a strong peak at $66.8\pm 0.4$\,d (1-$\sigma$ uncertainty). This period is significantly longer than the one given by \citet{hu17a}. This difference is to our different approach of subtracting a linear trend from the data, as well as the longer dataset used in this work. 
 We then fold the XRT light-curve on this period and mark the times of flux maximum in Fig.~\ref{fig:lc_uvot_xrt} based on the resulting profile. Note that this way we identify the data taken in December 2014 ($t=1600$) as the falling flank of the X-ray variability, i.e., the time of maximum flux occurred just before the XRT observation took place, while \citet{hu17a} assumed that the maximum flux occurred during the XRT observations. 

The peak in the Lomb-Scargle periodogram and the epoch folding result is relatively broad, indicating that the X-ray period might not be perfectly stable. This might indicate that the X-ray period is actually only quasi-periodic or that it varies on a super-orbital time-scale \citep[see, e.g.,][]{middleton18a}. However, for the sake of simplicity we will continue to refer to it as the X-ray period for the remainder of this paper.

\subsection{Pulsations}
\label{susec:puls}

\begin{table*}
\caption{Pulse period measurements.  New data are presented below the horizontal line. For clarity, we also list the epoch labels for the archival data given in \citet{p13} and \citet{walton18a}.The last columnd gives the pulsed fraction (PF). In the mission column ``X'' denotes \xmm observations and ``N'' denotes \nustar data.}
\label{tab:perevol}
\centering
\begin{tabular}{llllllll}
\hline\hline
Mission  & Epoch & ObsID & Date [MJD]  & P [ms] $-$415.0\,ms & $\dot P$ [$10^{-10}$\,s\,s$^{-1}$] & Flux\tablefootmark{a}  & PF [\%] \\\hline
X & X2/2013 & 0693760401  & 56621.21 & $4.712\pm0.008$ & $0.2^{+3.4}_{-2.8}$ & $7.03^{+0.28}_{-0.27}$ & $20.2\pm2.3$ \\
X & X3/2014 & 0748390901  & 57002.00 & $3.390^{+0.007}_{-0.008}$ & $-0.5^{+3.0}_{-2.5}$ & $19.5^{+0.6}_{-0.5}$ & $20.5\pm1.7$ \\
X & XN1/2016 & 0781800101  & 57528.58 & $1.951^{+0.008}_{-0.007}$ & $0.1^{+2.6}_{-2.9}$ & $36.2\pm0.6$ & $15.0\pm1.3$ \\
N & XN1/2016 & 80201010002  & 57528.18 & $1.9515^{+0.0016}_{-0.0019}$ & $-0.04^{+0.19}_{-0.17}$ & $36.2\pm0.6$ & $26.7\pm1.9$ \\
\hline
X & 2017A\tablefootmark{b}  & 0804670201  & 57886.17 & $0.932^{+0.018}_{-0.009}$ & $5^{+5}_{-14}$ & $17.1\pm0.8$ & $13\pm5$ \\
X & 2017B & 0804670301  & 57893.66 & $0.864^{+0.009}_{-0.006}$ & $-1.1^{+1.6}_{-3.2}$ & $15.7\pm0.4$ & $9.5\pm1.7$ \\
N & 2017B & 30302005002  & 57892.71 & $0.8755\pm0.0020$ & $-1.39^{+0.27}_{-0.22}$ & $15.7\pm0.4$ & $18.6\pm2.8$ \\
X & 2017C & 0804670401  & 57904.90 & $0.724\pm0.010$ & $-2\pm6$ & $31.7\pm0.8$ & $9.4\pm1.7$ \\
X & 2017D & 0804670501  & 57916.10 & $0.649^{+0.016}_{-0.025}$ & $2^{+13}_{-9}$ & $39.2^{+0.9}_{-0.8}$ & $7.6\pm1.5$ \\
X & 2017E & 0804670601  & 57924.11 & $0.669^{+0.008}_{-0.019}$ & $-6^{+12}_{-5}$ & $37.0\pm0.8$ & $9.5\pm1.6$ \\
N & 2017F & 30302015002  & 57933.93 & $0.7050^{+0.0024}_{-0.0017}$ & $0.65^{+0.22}_{-0.36}$ & $34.0\pm0.6$ & $15.4\pm2.1$ \\
N & 2017G & 30302015004  & 57942.93 & $0.7409^{+0.0023}_{-0.0011}$ & $0.13^{+0.16}_{-0.25}$ & $26.4\pm0.5$ & $14.2\pm2.4$ \\
N & 2017H & 90301326002  & 58057.58 & $0.2284^{+0.0035}_{-0.0030}$ & $0.4\pm0.7$ & $45.4\pm0.8$ & $12.5\pm2.1$ \\
X & 2017I & 0804670701  & 58083.00 & $0.214^{+0.007}_{-0.006}$ & $-1.0^{+2.4}_{-2.6}$ & $24.3\pm0.5$ & $13.8\pm1.4$ \\
N & 2017I & 30302005004  & 58082.95 & $0.2153^{+0.0018}_{-0.0024}$ & $-1.23^{+0.31}_{-0.24}$ & $24.3\pm0.5$ & $18.2\pm2.3$ \\
\end{tabular}
\tablefoot{
\tablefoottext{a}{Flux in $10^{-13}$\,erg\,s$^{-1}$\,cm$^{-2}$ between 3--10\,keV}
\tablefoottext{b}{Pulsations were not significantly detected in epoch 2017A, we here list the most prominent period found in the data.}
}
\end{table*}

To determine the pulse period $P$ we use all available \xmm and \nustar data (see Table~\ref{tab:perevol}), apart from the \xmm observations in 2012, which were taken during a very low flux state \citep{p13}. All \swift/XRT data are too short to identify the pulse period, given the low average count-rate of P13.
For \xmm we used EPIC-pn barycentered light-curves extracted on the fastest time-resolution of the detector of 73.4\,ms and between energies of 0.3--10\,keV.  For \nustar we used bary-centered and source-filtered event files (time-tagged with an intrinsic time resolution of 2\,$\mu$s), with energies between 3--79\,keV. To increase the signal-to-noise ratio (\snr), we combined the events of FPMA and FPMB. 

 As  a first step, we calculated a simple power-spectrum based on the fast Fourier transformation for each dataset. However, no coherent pulsations were significantly detected in any observation in 2017 using these power-spectra. This is contrary to the discovery of the pulsations, which showed up very clearly in the \xmm data of 2012, 2013, and 2014 \citep{p13, israel17b}. The non-detections are probably due the overall lower pulsed fractions and shorter exposure times of the new observations compared to previous observations.

In a next step we applied the epoch-folding technique, which is more sensitive to non-sinusoidal profiles. 
To narrow down our search range, we estimated the pulse period by extrapolating the secular spin-up as measured in our previous \xmm and \nustar observations \citep[$\dot P_\text{sec} \approx 3.5\times10^{-11}$\,s\,s$^{-1}$;][]{p13}. We searched  around that period in a range of $\pm 0.1$\,ms. In all observations besides 2017A this led to a clear detection of the pulse period. 

\begin{figure}
\includegraphics[width=0.95\columnwidth]{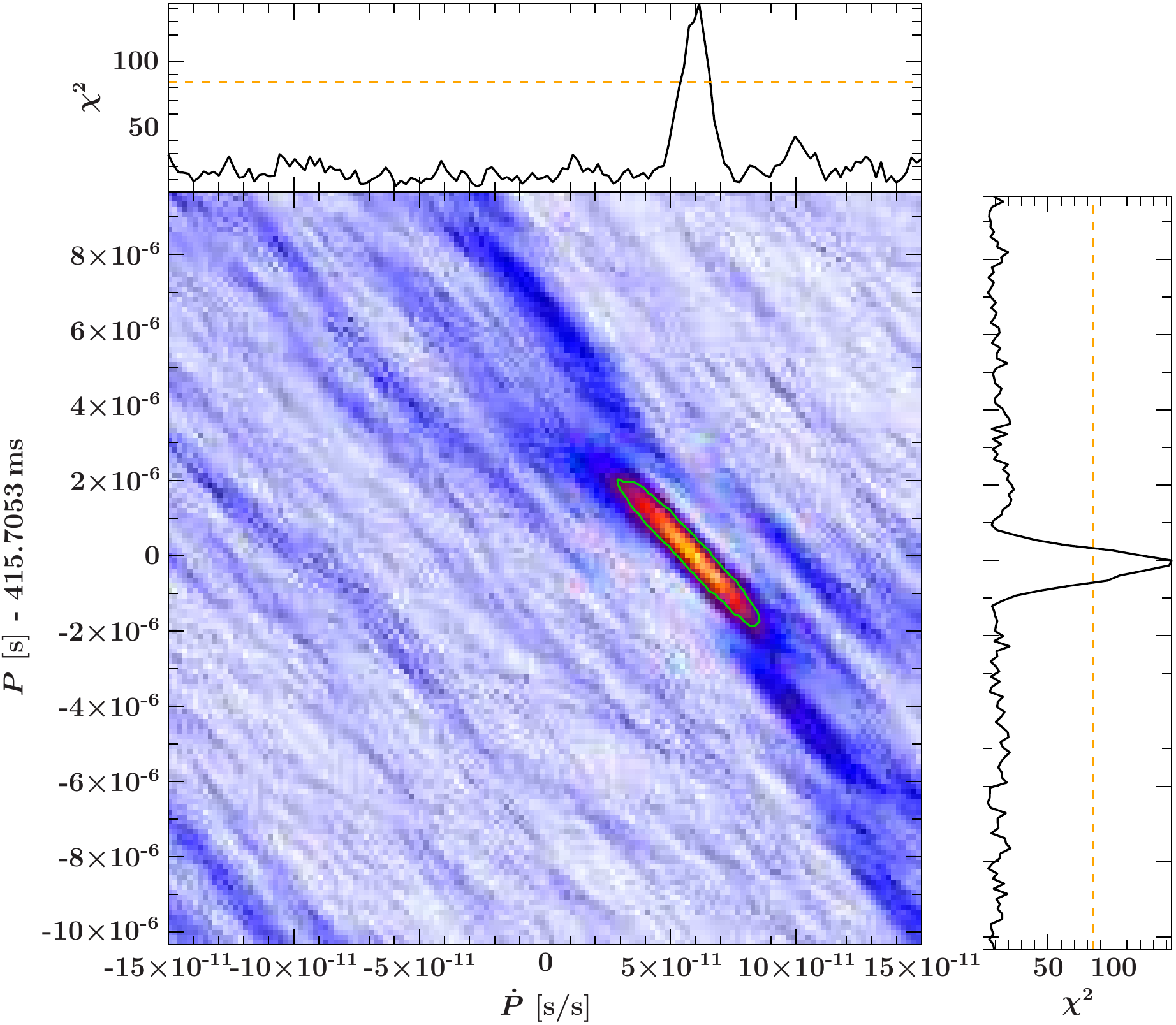}
\caption{Example of the epoch-folding in $P$-$\dot P$-space for \nustar observation 2017F. Color coded is the $\chi^2$ value at each $P$-$\dot P$ pair. The graphs on top and on the right-hand show the cut along the respective dimension for the most significant measurement at $P=415.7050$\,ms and $\dot P = 6.5\times10^{-11}$s\,s$^{-1}$. The orange dashed line show the 99.99\% significance line taking the number of trial periods into account.
}
\label{fig:epppdot}
\end{figure}

To refine the determination of $P$ and its derivative $\dot P$, we then use an accelerated epoch-folding search, i.e., searching a 2D-grid in $P$-$\dot P$ space. We typically used a grid size of 200$\times$200 points, centred on the best period found in the standard epoch folding, and on zero for $\dot P$, with a search range of $\pm 0.05$\,ms and $\pm10^{-9}$\,s\,s$^{-1}$, respectively. As in \citet{p13}, we determine the uncertainties on $P$ and $\dot P$ from the full-width half-maximum (FWHM) contour of the $\chi^2$ distribution. We show a typical example of this search in Fig.~\ref{fig:epppdot}.
We detect a period in each observation with a confidence larger than 99.99\% (apart from epoch 2017A, see below).  We calculated the significance based on a $\chi^2$ distribution with 11 degrees of freedom (i.e., the number of phase bins in each pulse profile minus one, \citealt{davies90a}) and took the number of trial periods (200 in $P$ and $\dot P$ each, i.e., a total of 40,000) into account.

Table~\ref{tab:perevol} gives the measured $P$ and $\dot P$ values in all observations. The first four measurements were published by \citet{p13} and the values for epochs X2/2013 and X3/2014 are also in good agreement with the ones found by \citet{israel17a}.
For the campaigns during which \xmm and \nustar observed quasi-simultaneously (epochs 2016, 2017B, and 2017I), we performed independent searches in \xmm and \nustar. While the values are fully consistent with each other, the \nustar observations provide the more  precise measurement due to their longer duration. In the rest of the analysis we will only use the \nustar value for those epochs.

Observation 2017A did not lead to a significant detection (at the 1-$\sigma$ limit) of the pulse period. However, we find a peak in the epoch-folding result around 415\,ms when selecting a window from between 5\,ks--15\,ks after the start of the observation. The rest of the observation is plagued by strong background flaring, which likely dilutes the pulsed signal. While this detection is still not significant by itself, we report it as well in Table~\ref{tab:perevol}, but do not use it in our further analysis. 

\subsection{Pulse profiles}

\begin{figure}
\includegraphics[width=0.95\columnwidth]{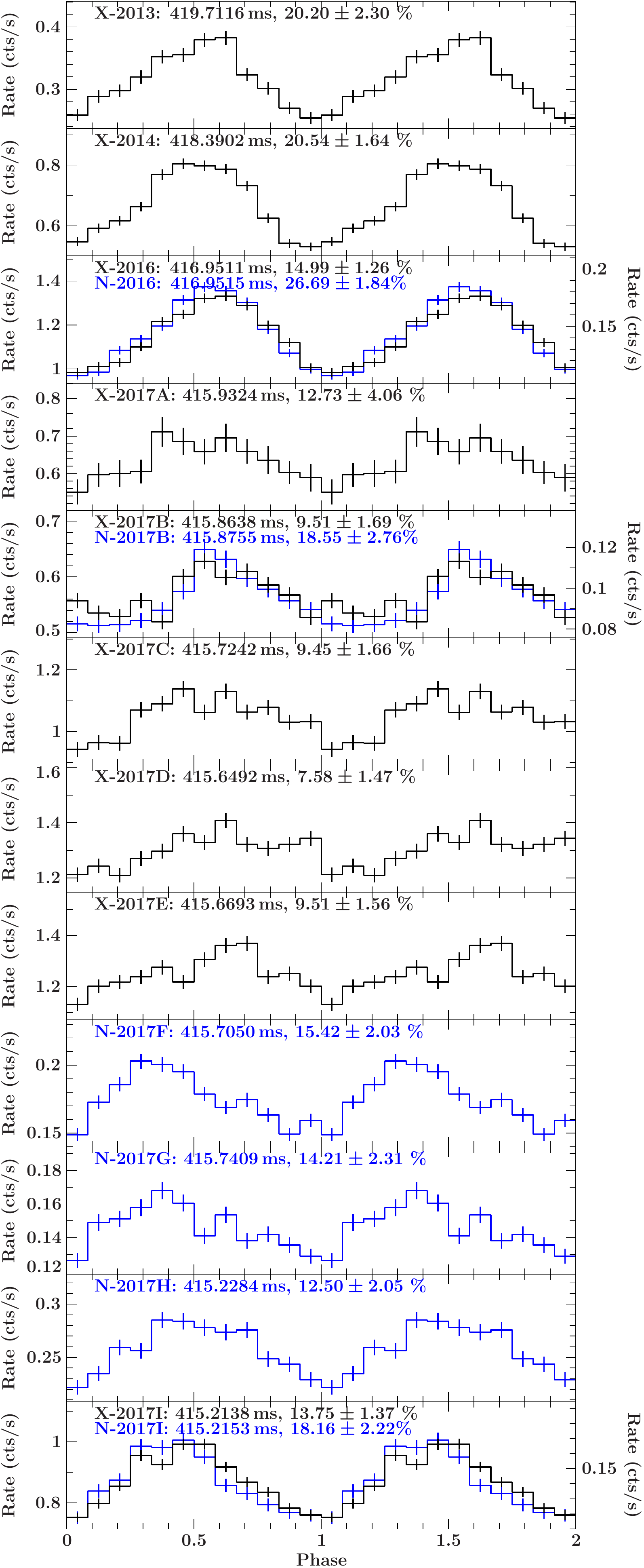}
\caption{Pulse profiles of all available \xmm and \nustar observations, folded on the respective measured pulse-period. The profiles are repeated once for clarity and phase-aligned by hand to have the  minimum at phase 0. 
For epochs where \xmm and \nustar observations where taken simultaneously, \nustar is shown in blue using the right-hand $y$-scale. Each panel is labeled with the epoch name, pulse period, and measured pulsed fraction in percent. For details see text.
}
\label{fig:allpp}
\end{figure}

To investigate the behavior of the pulse profile and the pulsed fraction as function of time, we folded all available data on their respective pulse period (Fig.~\ref{fig:allpp}). The data were binned into 12 phase-bins. We  calculated the pulsed fraction PF as
\[
\text{PF}_i=\frac{\max(p_i)-\min(p_i)}{\max(p_i)+\min(p_i)}
\]
where $p_i$ is the pulse profile of the respective observation.

As can be clearly seen in Fig.~\ref{fig:allpp}, the pulsed fraction during our 2017 campaign is on average much smaller than during the first three observations presented by \citet{p13}. Furthermore, the pulse profiles seem to be less sinusoidal, even when the pulsed fraction is relatively high, i.e., in epoch 2017B which shows a  narrower peak in both \xmm and \nustar.

\subsection{Determination of the orbit}
\label{susec:orbderm}

\begin{figure*}
\includegraphics[width=0.95\textwidth]{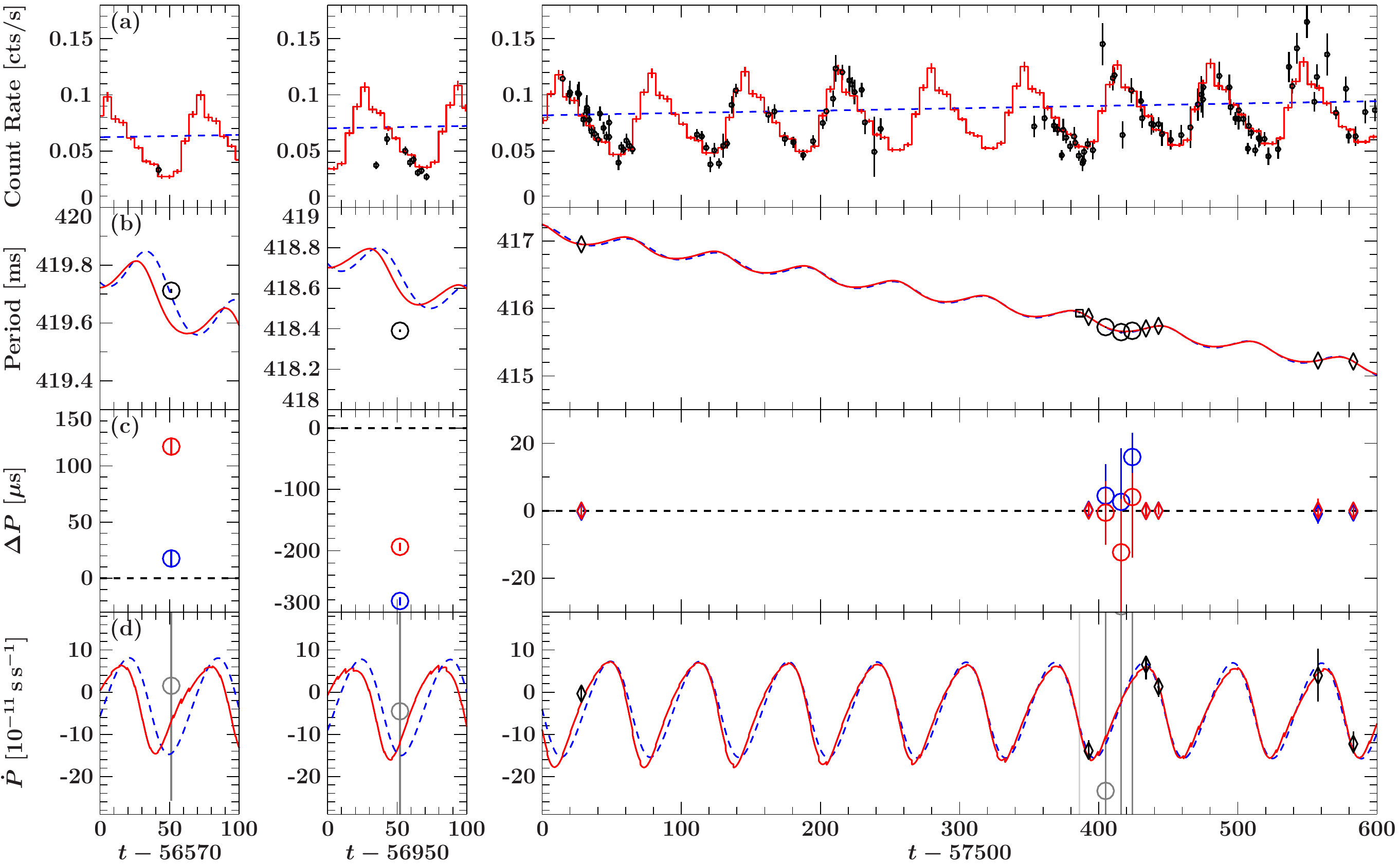}
\caption{\textit{(a)} \swift/XRT monitoring light-curve in the 0.3--10\,keV energy band.  The three columns focus on epochs 2013, 2014, and 2016--2017. The red curve shows the light-curve folded on a 66.9\,d period plus a linear brightening trend, the blue dashed line only the linear trend.  \textit{(b)} Measured values of the pulse period as a function of time.  Diamonds indicate values taken with \nustar, circles values taken with \xmm.  The small square shows the period measured in observation 2017A with \xmm, which did not provide a significant detection by itself. The red line is the best-fit model using the pulse profile as input, the blue dashed line a linear trend only. The model takes the secular spin-up as function of intrinsic luminosity and the orbital Doppler motion into account. For details see text. \textit{(c)} Residuals as data-minus-model for both models (red: pulse profile input, blue: linear trend only). \textit{(d)} Measured and predicted instantaneous change of the period ($\dot P$). The \xmm data are shown with grey circles and are not constraining due to the shorter exposure time of the \xmm observations.
Notice that the models were fitted without taking $\dot P$ into account. The good agreement between data and model is therefore an independent confirmation of the orbital ephemeris. The time axis is given in MJD.
}
\label{fig:orbit}
\end{figure*}

As can be seen in Table~\ref{tab:perevol} and Fig.~\ref{fig:orbit}(b), the pulse periods around MJD\,57900 show an apparent spin-down (between epochs 2017D to 2017G). This spin-down is a clear indicator of a strong Doppler effect due to the orbital motion, as a torque reversal on these time-scales is unlikely. We therefore fit an orbital model to the data between 2016--2017. Note that we do not use data from observation 2017A, as the period could not be independently found there and only use the \nustar values from epochs 2016, 2017B, and 2017I given their much higher precision compared to \xmm. 
We initially do not use older data, as over that long stretch of time a variable accretion torque is likely. 

The free parameters in the model are the orbital period, $P_\text{orb}$, the intrinsic pulse period, $P_\text{spin}$, at a pre-defined time, $T_0$, (which we fix at MJD~57530.00), the projected semi-major axis, $a\sin i$, the eccentricity, $e$, the time of periastron, $\tau$, and the argument of periastron, $\omega$. Additionally, we allow for an ad-hoc, constant spin-up parameter $\xi$. This model therefore has seven free parameters, compared to nine pulse period measurements, i.e., we only have two degrees of freedom. We do not find a good description of the data in terms of $\chi^2$ ($\chi^2$ = 4.9 for 2 dof), however, the secular spin-up trend and the indication for an orbital Doppler shift in 2017 is  well captured. We find an orbital period around 63.7\,d and a very small eccentricity ($e\approx0.05$). All parameters are given in Table~\ref{tab:orbfit}.

\begin{table*}
\caption{Best-fit orbital parameters for different input light-curves. The first column assumes no input light-curve, and therefore a constant spin-up $\xi$. The last column shows the results from the \texttt{emcee} run using the linear trend input light-curve.}\label{tab:orbfit}
\centering
\begin{tabular}{lllll}
\hline\hline
 Parameter & No Input  & Periodic Var. & Linear Trend &  Linear T. w/ MCMC   \\\hline
 $ P_\text{spin}~[\text{ms}]\tablefootmark{a}$ & $417.067^{+0.004}_{-0.006}$ & $417.027^{+0.007}_{-0.006}$ & $417.032^{+0.005}_{-0.004}$ & $417.032^{+0.013}_{-0.019}$ \\
 $\xi$  or  $b~[\text{s\,s}^{-1}]$\tablefootmark{b} & $\left(-4.034^{+0.009}_{-0.022}\right)\times10^{-11}$ & $\left(3.931^{+0.011}_{-0.029}\right)\times10^{-11}$ & $\left(3.648\pm0.009\right)\times10^{-11}$ & $\left(3.65^{+0.05}_{-0.06}\right)\times10^{-11}$ \\
 $ a\sin i~[\text{lt-s}]$ & $\left(2.61\pm0.08\right)\times10^{2}$ & $\left(2.01\pm0.08\right)\times10^{2}$ & $\left(2.09\pm0.08\right)\times10^{2}$ & $\left(2.09^{+0.18}_{-0.19}\right)\times10^{2}$ \\
 $ P_\text{orb}~[\text{d}]$ & $63.732\pm0.016$ & $64.226^{+0.018}_{-0.020}$ & $63.879^{+0.019}_{-0.013}$ & $63.9^{+0.6}_{-0.5}$ \\
 $ \tau$~[MJD] & $56574^{+73}_{-5}$ & $56668^{+10}_{-18}$ & $56672^{+24}_{-32}$ & $56669^{+26}_{-21}$ \\
 $ e$ & $0.05\pm0.04$ & $0.11^{+0.05}_{-0.06}$ & $0.05\pm0.05$ & $0.05^{+0.09}_{-0.05}$ \\
 $ \omega~[\text{deg}]$ & $\left(1.697^{+0.019}_{-0.007}\right)\times10^{2}$ & $38.1\pm2.3$ & $17.4\pm2.2$ & $-10^{+100}_{-110}$ \\
$\chi^2/\text{d.o.f.}$   & 9.89/2& 0.12/2& 0.51/2\\\hline
\end{tabular}
\tablefoot{
\tablefoottext{a}{at MJD~57530.00}
\tablefoottext{b}{spin-up for 0.08\,cts\,s$^{-1}$ in XRT}
}

\end{table*}

The observed pulse periods will also be influenced by the transfer of angular momentum onto the neutron star from the accreted matter. Following the theory by \citet{ghosh79a, ghosh79b}, the observed pulsed period change is proportional to  $P_\text{spin}L^{6/7}$ in the disk accretion case, where $L$ is the intrinsic luminosity of the system. The exact proportionality factor depends on different factors, like the magnetic field strength and the way the accretion disk couples to the magnetosphere. For simplicity, we describe it with a simple spin-up parameter $b$, for which we can fit directly 
\citep{marcu-cheatham15a, Bissinger2016}. The theory is based on a sub-critical, geometrically thin accretion disk, and we therefore caution that the coupling might be significantly different in the super-Eddington case. Nonetheless, this approach should give us a good indication about the transfer of angular momentum in the system.

We use the data from the \swift/XRT monitoring as a tracer of the intrinsic luminosity (Figure~\ref{fig:orbit}\textit{(a)}).
As the \swift/XRT monitoring has relatively large gaps, however, we do not use the data directly, but  interpolate the light-curve folded on the 66.9\,d X-ray period plus a linear brightening trend (see Sect.~\ref{susec:xrays}). This interpolation removes short term variability, but provides a good description of the overall behaviour of the X-ray light-curve (see Fig.~\ref{fig:orbit}). We thereby imply that the X-rays behaved in a similar way during the gaps of the XRT monitoring. As the spin-up depends on the luminosity, we give the proportionality factor $b$ in units of $\dot P$ at an XRT count-rate of $0.08$\,cts\,s$^{-1}$, which corresponds roughly to a luminosity of $L_0= 3.8\times10^{39} \ergcms$ (for a distance of 3.6\,Mpc).

With this approach we find a very good description of the pulse period evolution between 2016--2017 (Figure~\ref{fig:orbit}). The best-fit orbital period is determined to be $P_\text{orb}=64.226^{+0.018}_{-0.020}$\,d, and again the eccentricity is relatively small, $e=0.11^{+0.05}_{-0.06}$. All parameters are summarized in Table~\ref{tab:orbfit}. 

We find that the orbital period is very close to the period we found in the optical and UV data. This indicates that the optical photometry indeed varies mainly due to orbital effects, as discussed by \citet{motch14a}. These authors assume that the variability is due to the X-ray illuminated side of the optical companion rotating in and out of view. 

On the other hand, the timing solution clearly shows that the X-ray period is not the orbital period, but is significantly longer. It is therefore likely that the X-ray period is driven by a super-orbital period, on which, for example, the accretion disk or the neutron star could be precessing. 
As postulated by \citet{motch14a} and \citet{hu17a} it is also possible that the measured X-ray period in fact is a beat period between the orbital period and the super-orbital period, which would imply a super-orbital period of the order of 1500\,d. In any case, it seems plausible that the observed 66\,d X-ray variability is not a direct tracer of changes in accretion rate, but also strongly influenced by, e.g., the viewing angle onto the accretion disk \citep[similar to NGC\,5907 ULX,][]{ngc5907}.

However, if the X-rays are variable mainly due to geometrical effects, and not due to a real change in accretion rate, our approach of using the XRT light-curve as a tracer for the intrinsic luminosity is flawed. We therefore replace the input light-curve with just a simple linear brightening trend, based on the measured data. This approach also results in a very good description of the 2016--2017 pulse period data (Figure~\ref{fig:orbit}), and we find an orbital period of   $P_\text{orb}=63.984^{+0.019}_{-0.117}$\,d  and an eccentricity of $e=0.05\pm0.05$ (see Table~\ref{tab:orbfit}). This orbital period is in excellent agreement with the measured optical period.

We also test this model  including the 2013 and 2014 period measurements (Figure~\ref{fig:orbit}). However, we cannot achieve an acceptable fit with these epochs.
Extrapolating the model back to these times shows deviations up to 300\,$\mu$s in 2014, but a surprisingly good agreement for the linear trend model in 2013.  As we do not have good information about the luminosity between these measurements, different accretion torques might have been acting onto the neutron star during the long gaps of no X-ray monitoring. Changes of the luminosity of a factor $<2$ are enough to change the inferred pulse period by an amount similar to the deviations observed, making this hypothesis likely.

Finally, we compare the measured values of the instantaneous period change, $\dot P$, in each epoch to the ones predicted by the model. This information is not used during fitting, and therefore provides an independent test of the model predictions. As shown in Figure~\ref{fig:orbit}(d), the model agrees very well with the measured $\dot P$ values, however, only the \nustar measurements are constraining. We show the complete expected $\dot P$ value, i.e., the combination between secular spin-up and orbital motion, to be directly comparable with the measured values.

\subsubsection{Uncertainty estimation through Monte Carlo simulations}

As we are fitting a complex model to a relative sparse dataset, we run  Markov Chain Monte Carlo  (MCMC) simulations to investigate parameter degeneracies and obtain a more realistic estimate of the uncertainties.  We use the \texttt{emcee} method implementation in ISIS, which is based upon the parallel ``simple
 stretch'' method presented by \citet{foreman12a}. We run 200 walkers per free parameter for 5000 steps each. As input light-curve we use the simple linear trend, as the model that seems to describe the system's behavior most realistically. We evaluate the model for the 2016 and 2017 data only.

\begin{figure*}
\includegraphics[width=1.1\textwidth]{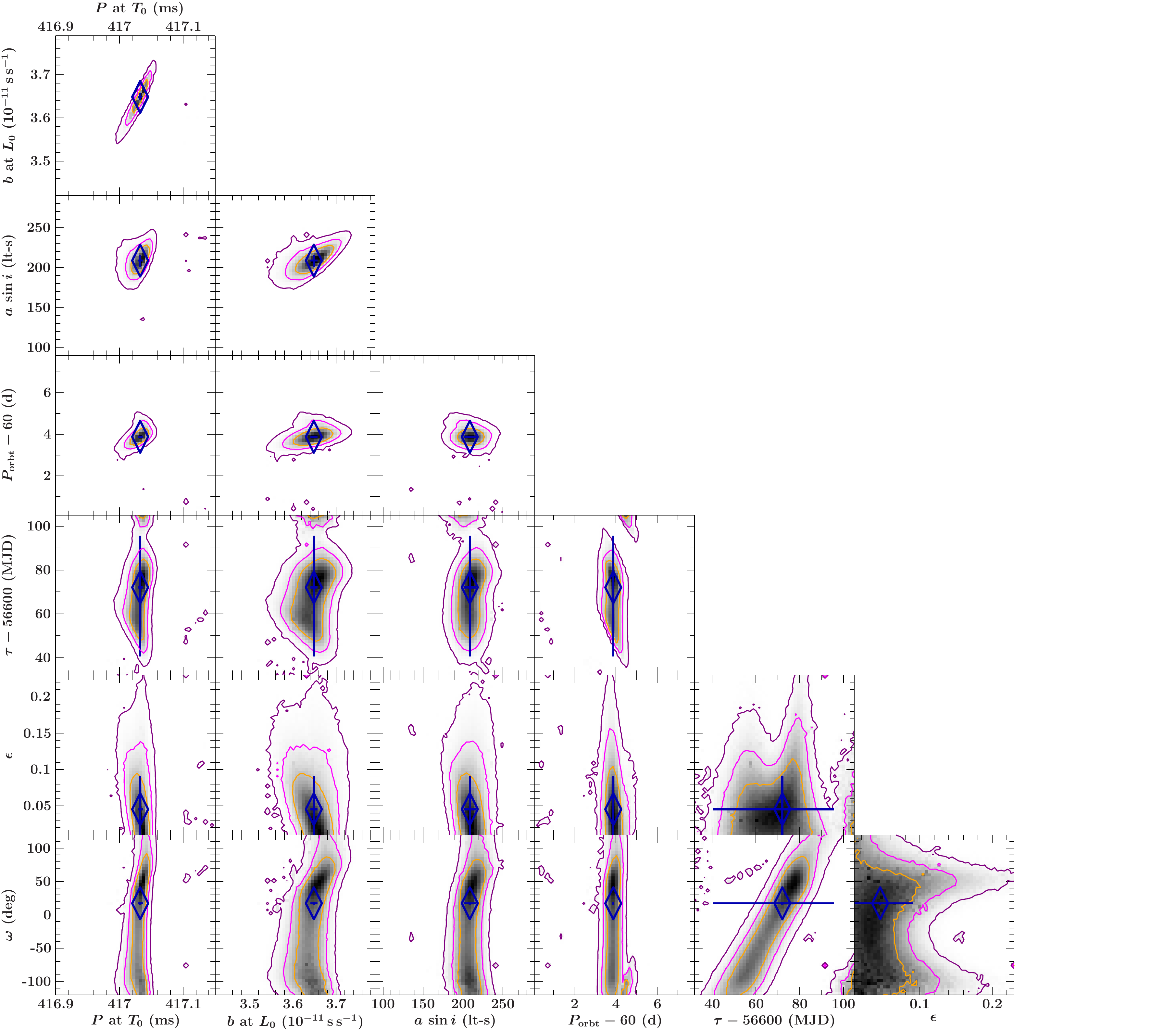}
\caption{Results of the MCMC run for the orbital determination, based on a linearly brightening input light-curve. The blue diamonds give the best-fit values and uncertainties from a standard fit. The contour lines give the 68\%, 90\%, and 99\%  confidence levels, i.e., number of walkers contained within these regions.}
\label{fig:emcee}
\end{figure*}

We analyse the results from the MCMC run after a burn-in period of 30\% (1500 steps). Figure~\ref{fig:emcee} shows the results for each parameter pair, together with the best-fit values from a standard $\chi^2$-minimisation fit. As can be seen in the Figure, no strong degeneracies are present, beside between the argument of periastron $\omega$ and the time of periastron passage $\tau$, which can be understood given that the eccentricity is so low. In a circular orbit, these two parameters would be perfectly degenerate within one orbital period. The orbital period is in particular very well determined, and periods longer than 65\,d are ruled out at the 99\% level.

 As degeneracies are small, we also give the estimated 90\% uncertainties from the 1D-distributions in Table~\ref{tab:orbfit}. These uncertainties are larger than the ones estimated from a simple $\chi^2$-fit,  which shows that the main driver of the uncertainties are systematics. In particular the small number of data-points we currently have makes a precise determination of all orbital parameters difficult. Nonetheless, we find that the measured orbital period still is significantly different to the X-ray flux period.

Last, we check that there are no acceptable solutions for orbital periods $<10$\,d, which would be more similar to the orbital periods proposed for M82\,X-2 and \ngculx. We run the same analysis, in particular an MCMC simulation with 200 walkers with 5000 steps each in the parameters space $0.5\,\text{d} \leq P_\text{orb} \leq 10\,\text{d}$. As expected, we do not find an acceptable solution, with a best-fit $\chi^2 > 10$ and no stable minimum in $\chi^2$-space. This result is in line with the expectations from the binary properties, where a minimal orbital period of $P_\text{orb} \approx 24$\,d is required for the neutron star to be outside the stellar photosphere for a typical B9Ia star.

\section{Discussion \&  Conclusion}
\label{sec:disc}

We have  presented a timing analysis of  an \xmm and \nustar campaign of the ultra-luminous pulsar NGC7793~P13, carried out in late 2017. 
Through the  variation of the pulse period, we determined an orbital period  between $\sim$63.7--64.2\,d, consistent with the optical photometric period. On top of the orbital variation, we observe a strong secular spin-up, which we relate to transfer of angular momentum from the accreted material. When using a simple linear brightening trend as function of time, and momentum transfer based on the theory by \citet{ghosh79a}, we find a very good description of the pulse period evolution from 2016--2017. We are unable to include older data in the model, as we lack the necessary information about the intrinsic X-ray flux.

As shown in Fig.~\ref{fig:orbit}, we also find a  good description of the pulse periods when assuming that the measured X-ray count-rate is a direct tracer of the X-ray luminosity. In that case, we find a slightly longer orbital period of 64.2\,d. The current available data does not allow us to distinguish between these two scenarios. Another observation campaign in 2019 may be able to clearly distinguish between the models, if the average X-ray flux  until then is well monitored. 

Our best-fit model with a linear brightening trend implies that the observed periodic X-ray variability is not dominated by changes in the accretion rate, but more likely by geometric effects from a precessing accretion disk. This explanation is in line with the fact that the X-ray period is significantly longer than the orbital period, around $P_\text{X}= 66.9$\,d and that we observe an almost circular orbit. 
A possible driver for the disk precession could be the Lense-Thirring effect \citep{middleton18a}.

It is  also possible that the X-ray period is the beat period of the orbital period with a very long super-orbital period. That implies a super-orbital period of around 1500\,d. Similar super-orbital periods have been suggested by \citet{motch14a} and \citet{hu17a} based on an observed phase jitter of the times of maximum light in the optical. If this interpretation is correct, it implies that the X-ray flux is also significantly influenced by the orbital period. This might be possible through the small, but non-negligible eccentricity we find, which could influence the accretion rate as function of orbital phase. Based on the simple model by \citet{brown84a}, variations of a factor of 2 in mass accretion rate over the orbit are certainly possible for an eccentricity of $e=0.1$, even for relatively large photospheres of the donor star.

However, the fact that the X-ray period is only a few percent longer than the optical period is reminiscent of optical super-hump periods observed in SU Ursae Majoris dwarf novae \citep{warner95a} which have also been observed in accreting black-hole systems \citep{odonoghue96a}. The super-hump is caused by a 3:1  resonance between the Keplerian velocity in the outer  accretion disk and the orbital period, resulting in tidal forces that produce a finite eccentricity as well as a nodal precession of the accretion disk \citep[e.g.,][]{whitehurst88a, lubow91a}. While the P13 system has an inverse mass ratio compared to typical dwarf novae (i.e., the  donor is significantly more massive than the accreting object), it is nonetheless extreme ($q=M_\text{NS}/M_\star \approx 0.08$). A similar resonance could therefore take place, observable as periodicity in the X-rays emitted by the accretion disk. Following the calculations by \citet{whitehurst91a}, we find that indeed a 4:1 resonance is possible within the tidal radius of the neutron star. To determine the exact super-hump period and strength hydrodynamic simulations are necessary, which are beyond the scope of this work.
Additionally, tidal forces due to a misalignment of the neutron star spin axis with the orbital axis could also induce a warp or precession of the accretion disk \citep{martin09a}.

Using our orbital solution and assuming a canonical mass of about 1.4\,\msun for the neutron star and between 18--23\,\msun for the B9I companion, we can solve the mass function for the inclination. We obtain a semi-major axis of  180--196\,\rsun, assuming $P_\text{orb}=64$\,d and $e=0.1$. With the measured $a \sin i$ of 208\,lt-s ($\approx$90\,\rsun) we find an inclination of around $30^\circ$. This is in agreement with the fact that we do not observe eclipses, i.e., we cannot be observing the system edge-on. 

With these orbital parameters we find a Roche-lobe radius at periastron of about 103--119\,\rsun.  On the other hand  for a temperature of 11$\pm$1\,kK and an absolute V-band magnitude of $M_\text{V} = -7.44$ we derive a radius of 96--125\,\rsun for the B9I donor star, with the method used by the genetic algorithm described by \citet{mokiem05a}.
 The star  therefore  subtends over about a 60$^\circ$ cone of the sky, as seen from the neutron star.  It is very likely that the star fills its Roche Lobe even with the small eccentricity implied, allowing for the high accretion rate necessary to power the observed X-ray luminosity of the system. However, we note that the stellar parameters are not very well determined, for example the absolute magnitude could be contaminated by optical emission from the accretion disk. It is therefore  possible that the stellar radius is smaller, however, the high observed luminosity strongly argues in favor of Roche-Lobe accretion.

P13 shows a hard ULX spectrum \citep{walton18a}, consistent with the ``hard ultraluminous state`` as defined by \citet{sutton13a}.  In the ``ULX unification model'' \citep{sutton13a, middleton15a} such a hard spectrum implies that we observe the source at relatively low inclination angles, down the evacuated cone surrounded by the super-Eddington disk and its wind \citep{pinto16a}. If the accretion disk and neutron star inclination is aligned with the orbital inclination, our measurements would therefore be in agreement with this model, assuming the cone has an opening angle of at least $30^\circ$. The opening angle (of, at least, the soft X-rays) would need to be larger to also illuminate the star and thereby explain the optical period.
However, we note that this model might not hold for accreting neutron stars with strong magnetic fields, and alternative models have been discussed \citep[see, e.g.][]{koliopanos17a}.

To obtain a significant effect of geometric beaming, which would help to explain the apparent super-Eddington luminosity of P13 without the need to evoke extreme accretion rates, \citet{dauser17a} find that very narrow cones are necessary. To model the super-orbital variability of \ngculx \citep{walton16b}, they require half-opening angles of $\ll10^\circ$ and for half-opening angles of 25$^\circ$ or larger the amplification factor rapidly drops to around 1. Note, however, that the model of \citet{dauser17a} is based on a rather simplified geometry and does not take the emission geometry of the accretion column into account.

All our orbital solutions require a very small eccentricity $e\leq0.15$. \citet{motch14a} implied a significantly larger eccentricity between 0.3--0.4, based on their modelling of the optical light-curve. In particular, they argue that the sharp optical maxima imply an eccentric orbit, if they are due to the X-ray illuminated side of the star. However, the accretion geometry of the system is currently unknown, and some of the optical flux could also be contributed by an accretion stream or hot spot on the super-Eddington accretion disk.

For P13,  pulsations have so far been seen in every observation that provides a high enough \snr. Two of the other pulsating ULXs have shown pulse drop outs, despite no significant change in flux \citep{israel17a, bachetti14a}. Here we have shown, however, that P13 also shows strong variability in its pulsed fraction, with values as low as 8\% during epoch 2017D.  Additionally the measured pulse profile shows considerable variability, deviating from the previously observed sinusoidal shape \citep{p13}. This variability shows that the accretion and emission geometry changes on relatively short time-scales. A detailed analysis of the pulsed spectrum is beyond the scope of this paper, however, and will be presented in a forthcoming work (Walton et al., in prep.).

\begin{acknowledgements}

We would like the thank the referee for the very helpful comments that helped to improve the manuscript.
We thank M.~K\"uhnel  and M.~Nowak for the useful discussions.
DJW and MJM  acknowledge support from STFC Ernest Rutherford fellowships.
This research has made
use of data obtained with \nustar, a project led by Caltech, funded by NASA and
managed by NASA/JPL, and has utilized the \texttt{nustardas} software package, jointly
developed by the ASDC (Italy) and Caltech (USA). This research has also made
use of data obtained with \xmm, an ESA science mission with instruments and 
contributions directly funded by ESA Member States. This work made
use of data supplied by the UK Swift Science Data Centre at
the University of Leicester, and also made use of the XRT
Data Analysis Software (XRTDAS) developed under the responsibility
of the ASI Science Data Center (ASDC), Italy.
This research has made use of a collection of ISIS functions (ISISscripts) provided by ECAP/Remeis observatory and MIT (\url{http://www.sternwarte.uni-erlangen.de/isis/})

\end{acknowledgements}

\end{document}